# Non-Prefered Reference Frames and Anomalous Earth Flybys

Walter Petry
Mathematisches Institut der Universitaet Duesseldorf, D-40225 Duesseldorf
E-mail: wpetry@meduse.de
petryw@uni-duesseldorf.de

Abstract: Let us consider a reference frame $\Sigma'$ in which the pseudo-Euclidean geometry holds. Einstein assumed that the principle of special relativity is valid, i.e. the reference frame of any uniformly moving observer is also described by the pseudo-Euclidean geometry. The transformation formulae from one reference frame to the other one are therefore given by the well-known Lorentz-transformations. But Einstein's assumption contradicts the observed dipole anisotropy of the cosmic microwave background (CMB) in the universe. The transformation formulae of the prefered reference frame $\Sigma'$ in which the pseudo-Euclidean geometry is valid to a uniformly moving observer in a non-prefered reference frame $\Sigma$ are stated. The geomerty in $\Sigma$ is anisotropic. The Doppler shift of objects moving in a non-prefered reference frame is calculated. This result is applied to spacecrafts which fly near the Earth. The observed anomalous frequency shift of several spacecrafts during near Earth flybys does not arise in the non-prefered reference frame $\Sigma$.

## 1. Introduction

Let us consider a reference frame $\Sigma'$ in which the pseudo-Euclidean geometry holds. The principle of Einstein's special relativity theory states that the pseudo-Euclidean geometry is valid for any uniformly moving reference frame. The transformation formulae leaving the pseudo-Euclidean geometry invariant are given by the well-known Lorentz-transformations. But the principle of special relativity contradicts the observed dipole anisotropy of the CMB in the universe. Therefore, the Lorentz-transformations which do not change the pseudo-Euclidean geometry are interpreted as transformations of an event at rest in $\Sigma'$ to the same event moving uniformly in the frame $\Sigma'$. The geometry of a non-prefered reference frame $\Sigma$ moving uniformly relative to the prefered frame $\Sigma'$ is anisotropic. The transformation formulae from the frame $\Sigma'$ to the frame $\Sigma$ and conversely are stated. Analogous formulae to the Lorentz-transfomations in the reference frame $\Sigma'$ are given in the frame $\Sigma$, i.e. the geometry in $\Sigma$ is unchanged by these transformation formulae. Hence, knowing an event in $\Sigma$ at rest one can receive the same event which moves uniformly in $\Sigma$. Although the light-velocity in $\Sigma$ is anisotropic the theory gives the null result of the Michelson-Morley experiment.These results and many others are considered by the author in paper [1] . There exist other formulae which transform the prefered reference frame $\Sigma'$ to the non-prefered reference frame $\Sigma$ which are e.g. given by Marinov [2]. Many authors have also studied these transformation formulae from $\Sigma'$ to a non-prefered frame $\Sigma$.

The Doppler frequency of a moving object in the non-prefered reference frame $\Sigma$ is derived by the use of the transformation formulae which do not change the geometry in $\Sigma$. This result is applied to spacecrafts which flyby near the Earth. In the non-prefered reference frame the anomalous Doppler frequency shift of spacecrafts during the Earth flyby does not arise, i.e. the anomalous Earth flyby can be exlpained. All the considerations about the anomalous Earth flyby use the prefered reference frame.

It is worth to mention that the observed Pioneer anomaly is received by considering the Pioneers in the solar system without universe whereas this anomaly does not arise if the universe is included ( see [3]).

## 2. Prefered and Non-Prefered Reference Frames

In this section we follow along the lines of paper [1]. The pseudo-Euclidean geometry holds in the prefered reference frame $\Sigma'$. All the quantities in this frame are denoted with a prime. Let $(x^{i'})$ be the space-time vector where $x' = (x^{1'}, x^{2'}, x^{3'})$ are the Cartesian coordinates and $x^{4'} = ct'$. Then, the line-element has the form

$$(ds')^2 = -\eta_{ij}' dx^{i'} dx^{j'} \tag{2.1}$$

with $(\eta_{ij}') = diag(1,1,1,-1)$. The tensor $(\eta^{ij'})$ is defined by

$$\eta_{ik}' \eta^{kj'} = \delta_i^{\ j}.$$

Let $(\cdot,\cdot)$ denote the scalar product in $R^3$ and $|\cdot|$ the induced Euclidean norm. Let $w' = (w^{1'}, w^{2'}, w^{3'})$ be a constant velocity vector and put $\gamma = \left(1 - \left|\frac{w'}{c}\right|^2\right)^{-1/2}$ then the Lorentz transformations have the form

$$\tilde{x}^{i\,\prime} = x^{i\,\prime} + (\gamma - 1)\frac{(x',w')}{|w'|^2}w^{i\,\prime} + \gamma x^{4\,\prime}\frac{w^{i\,\prime}}{c}$$

$$\tilde{x}^{4\,\prime} = \gamma\left(x^{4\,\prime} + \left(x',\frac{w'}{c}\right)\right). \tag{2.2}$$

The inverse formulae are given by

$$x^{i\,\prime} = \tilde{x}^{i\,\prime} + (\gamma - 1)\frac{(\tilde{x}',w')}{|w'|^2}w^{i\,\prime} - \gamma\tilde{x}^{4\,\prime}\frac{w^{i\,\prime}}{c}$$

$$x^{4\,\prime} = \gamma\left(\tilde{x}^{4\,\prime} - \left(\tilde{x},\frac{w'}{c}\right)\right). \tag{2.3}$$

It is well known that the Lorentz-transformations (2.2) resp. (2.3) do not change the pseudo-Euclidean geometry (2.1). Hence, an event at rest in $\Sigma'$ described by the coordinates $(x^{i\,\prime})$ is transformed to the same event in $\Sigma'$ with the coordinates $(\tilde{x}^{i\,\prime})$ which moves with constant velocity $w'$ relative to the first one. Let us now consider a frame $\Sigma$ uniformly moving with constant velocity $-v' = (-v^{1\,\prime}, -v^{2\,\prime}, -v^{3\,\prime})$ relative to $\Sigma'$. Let us assume that the transformation formulae from $\Sigma'$ to $\Sigma$ are given by

$$x^{i\,\prime} = x^i \quad (i = 1,2,3), \qquad x^{4\,\prime} = x^4 + \left(x,\frac{v'}{c}\right) \tag{2.4}$$

where $(x^i)$ is the four-vector of space-time in $\Sigma$. The inverse formulae have the form

$$x^i = x^{i\,\prime} \quad (i = 1,2,3), \qquad x^4 = x^{4\,\prime} - \left(x',\frac{v'}{c}\right). \tag{2.5}$$

Then the transformation formulae for tensors together with (2.4) resp.(2.5) give the line-element in $\Sigma$

$$(ds)^2 = -\eta_{ij}dx^i dx^j \tag{2.6}$$

with

$$\eta_{ij} = \delta_{ij} - \frac{v^{i\,\prime}}{c}\frac{v^{j\,\prime}}{c} \quad (i, j = 1,2,3)$$

$$\eta_{i4} = \eta_{4i} = -\frac{v^{i\,\prime}}{c} \quad (i = 1,2,3)$$

$$\eta_{44} = -1 \tag{2.7}$$

and

$$\eta^{ij} = \delta^{ij} \quad (i, j = 1,2,3)$$

$$\eta^{i4} = \eta^{4i} = -\frac{v^{i\,\prime}}{c} \quad (i = 1,2,3)$$

$$\eta^{44} = -\left(1 - \left|\frac{v'}{c}\right|^2\right). \tag{2.8}$$

The relations (2.7) resp. (2.8) together with (2.6) imply an anisotropy in the reference frame $\Sigma$.

By the use of the transformations (2.4) of $\Sigma'$ to $\Sigma$ resp. the corresponding ones for moving objects

$$\tilde{x}^{i\,\prime} = \tilde{x}^i \quad (i = 1,2,3), \qquad \tilde{x}^{4\,\prime} = \tilde{x}^4 + \left(\tilde{x},\frac{v'}{c}\right) \tag{2.9}$$

and the Lorentz-transformations (2.2) resp. (2.3) we get the transformations leaving the line-element (2.6) with (2.7) in $\Sigma$ invariant. Hence, these transformations correspond to the Lorentz-transformations in $\Sigma'$, i.e. an event which moves with constant velocity in the frame $\Sigma$ can be calculated if the same event at rest in $\Sigma$ is known. The resulting transformations have the form

$$\tilde{x}^i = x^i + \gamma x^4 \frac{w^{i\prime}}{c} + (\gamma-1)\frac{(x,w')}{|w'|^2} w^{i\prime} + \gamma\left(x,\frac{v'}{c}\right)\frac{w^{i\prime}}{c} \quad (i=1,2,3)$$

$$\tilde{x}^4 = \gamma x^4\left(1-\left(\frac{w'}{c},\frac{v'}{c}\right)\right) + \left(\gamma-1-\gamma\left(\frac{w'}{c},\frac{v'}{c}\right)\right)\left(x,\frac{v'}{c}\right) + \left(\gamma-(\gamma-1)\frac{(w',v')}{|w'|^2}\right)\left(x,\frac{w'}{c}\right) \quad (2.10)$$

and

$$x^i = \tilde{x}^i - \gamma\tilde{x}^4\frac{w^{i\prime}}{c} + (\gamma-1)\frac{(\tilde{x},w')}{|w'|^2}w^{i\prime} - \gamma\left(\tilde{x},\frac{v'}{c}\right)\frac{w^{i\prime}}{c} \quad (i=1,2,3)$$

$$x^4 = \gamma\tilde{x}^4\left(1+\left(\frac{w'}{c},\frac{v'}{c}\right)\right) + \left(\gamma-1+\gamma\left(\frac{w'}{c},\frac{v'}{c}\right)\right)\left(\tilde{x},\frac{v'}{c}\right) - \left(\gamma+(\gamma-1)\frac{(w',v')}{|w'|^2}\right)\left(\tilde{x},\frac{w'}{c}\right) \quad (2.11)$$

The transformations of the velocities $v'$ and $w'$ in the frame $\Sigma'$ to velocities $v$ and $w$ in $\Sigma$ can be found in paper [1].

In the reference frame $\Sigma$ the frame $\Sigma'$ is descibed by the velocity $w' = v'$. The formulae (2.10) and (2.11) together with (2.5) give the transformations

$$\tilde{x}^i = x^{i\prime} + (\gamma-1)\frac{(x',v')}{|v'|^2}v^{i\prime} + \gamma x^{4\prime}\frac{v^{i\prime}}{c} \quad (i=1,2,3)$$

$$\tilde{x}^4 = \gamma^{-1}x^{4\prime} \quad (2.12)$$

from the frame $\Sigma'$ in the frame $\Sigma$ with $\gamma = \left(1-\left|\frac{v'}{c}\right|\right)^{-1/2}$. The inverse fomulae are

$$x^{i\prime} = \tilde{x}^i + (\gamma^{-1}-1)\frac{(\tilde{x},v')}{|v'|^2}v^{i\prime} - \gamma\tilde{x}^4\frac{v^{i\prime}}{c} \quad (i=1,2,3)$$

$$x^{4\prime} = \gamma\tilde{x}^4. \quad (2.13)$$

The formulae (2.12) resp. (2.13) which transform the frame $\Sigma'$ in the frame $\Sigma$ and conversely are studied by many authors (see e.g. [2] ) where non-prefered reference frames are studied, too

It is worth to mention that in paper [1] Maxwell's equations and equations of motion in the non-prefered reference frame $\Sigma$ are stated. Furthermore, several applications in the reference frame $\Sigma$ are given.

### 3. Earth Flyby Anomaly

Let $(J^i)$ denote the four-vector of the electrical current. Then, Maxwell's equations can be rewritten by the use of the Lorentz gauge (see paper [1])

$$\frac{\partial}{\partial x^i}\left(\eta^{ij}A_j\right) = 0 \quad (3.1)$$

in the form

$$\frac{\partial}{\partial x^j}\left(\eta^{jk}\frac{\partial A_i}{\partial x^k}\right) = \frac{4\pi}{c}\eta_{ij}J^j \quad (3.2)$$

where $(A_i)$ are the electro-magnetic potentials.

Let us now consider a plane wave in the reference frame $\Sigma$, i.e. $J^i = 0 \ (i=1,2,3,4)$. This wave is described in the form

$$A_i = A_{i0}\cos(k_j x^j) \quad (3.3)$$

where $A_{i0}$ and $k_j$ are constants. It follows by the equations (3.1) and (3.2)

$$\eta^{ij}k_i k_j = 0, \quad \eta^{ij}k_i A_{j0} = 0. \quad (3.4)$$

The first equation of (3.4) implies by the use of (2.8) with the vector $k = (k_1, k_2, k_3)$ the relation

$$|k|^2 - 2\left(\frac{v'}{c}, k\right)k_4 - \left(1 - \left|\frac{v'}{c}\right|^2\right)k_4^2 = 0. \tag{3.5}$$

Equation (3.5) gives

$$k_4 = |k|\left(\left(1 - \left|\frac{v'}{c}\right|^2 \sin^2(k;v')\right)^{1/2} - \left|\frac{v'}{c}\right|\cos(k;v')\right) / \left(1 - \left|\frac{v'}{c}\right|^2\right). \tag{3.6}$$

Here, $\sin(v';k)$ denotes the function $\sin$ of the angle between the vectors $v'$ and $k$. The second relation of (3.4) states a condition for the constants $A_{i0}$ which is not needed subsequently.

Let us now assume that a plane wave is emitted from an object moving in $\Sigma$ with the constant velocity $w' = (w^{1'}, w^{2'}, w^{3'})$. The corresponding wave four-vector is denoted by $\tilde{k} = (\tilde{k}_1, \tilde{k}_2, \tilde{k}_3, \tilde{k}_4)$. The well known transformation formulae of four-vectors

$$\tilde{k}_i = k_j \frac{\partial x^j}{\partial \tilde{x}^i} \tag{3.7}$$

give by the use of the transformation formulae (2.11) for moving objects in $\Sigma$ the fourth component

$$\tilde{k}_4 = -\gamma\left(k, \frac{w'}{c}\right) + \gamma k_4 \left(1 + \left(\frac{w'}{c}, \frac{v'}{c}\right)\right) = \gamma k_4\left(1 + \left(\frac{w'}{c}, \frac{v'}{c}\right)\right) - \gamma|k|\left|\frac{w'}{c}\right|\cos(k;w').$$

It follows by the use of relation (3.6)

$$\tilde{k}_4 = \gamma k_4\left(1 + \left(\frac{w'}{c}, \frac{v'}{c}\right) - \left|\frac{w'}{c}\right|\frac{\left(1 - \left|\frac{v'}{c}\right|^2\right)\cos(k;w')}{\left(1 - \left|\frac{v'}{c}\right|^2\sin^2(k;v')\right)^{1/2} - \left|\frac{v'}{c}\right|\cos(k;v')}\right) \tag{3.8}$$

We get by neglecting higher order terms of the velocities

$$\tilde{k}_4 \approx \gamma k_4\left(1 - \left|\frac{w'}{c}\right|\cos(k;w') + \left(\frac{w'}{c}, \frac{v'}{c}\right) - \left|\frac{w'}{c}\right|\left|\frac{v'}{c}\right|\cos(k;w')\cos(k;v')\right). \tag{3.9}$$

Let us introduce the frequency $\nu$ then (3.9) implies

$$\tilde{\nu} \approx \gamma\nu\left(1 - \left|\frac{w'}{c}\right|\cos(k;w')\right) + \nu\left(\frac{w'}{c}, \frac{v'}{c}\right) - \nu\left|\frac{w'}{c}\right|\left|\frac{v'}{c}\right|\cos(k;w')\cos(k;v'). \tag{3.10}$$

where $\nu$ is the wave frequency emitted by the object at rest in $\Sigma$ and $\tilde{\nu}$ is the received wave frequency emitted by the same object when it moves with the velocity $w'$.

The first expression in relation (3.10) is the frequency in the prefered reference frame whereas the second expression is the correction implied by the non-prefered reference frame.

In many papers the anomalies of several spacecrafts that fly past the Earth on approximately hyperbolic trajectories are given (see e.g. [4,5] and the cited literature therein ). The Doppler residuals, i.e. the observed frequency $\nu_{obs}$ minus the computed Doppler frequency during the Earth flyby of the spacecrafts show the same frequency behavior. The pre-perigee residual is nearly constant then there is a jump in the neighborhood of the nearest point to the Earth to a nearly constant post-perigee residual. This result is called "anomalous Earth flyby". Let us now appoly romula (3.10) implying

$$\frac{\nu_{obs} - \tilde{\nu}}{\nu} \approx \frac{\nu_{obs} - \nu\gamma\left(1 - \left|\frac{w'}{c}\right|\cos(k;w')\right)}{\nu} - \left|\frac{w'}{c}\right|\left|\frac{v'}{c}\right|(\cos(w';v') - \cos(k;w')\cos(k;v')). \tag{3.11}$$

The first expression of the right hand side gives the anomalous Earth flyby (all the authors use the prefered reference frame). The second expression gives a correction by virtue of the non-prefered frame of the Earth. It holds in the neighborhood of the nearest point to the Earth of the spacecraft

$$|\cos(k;w')| \ll 1. \tag{3.12}$$

The absolute value of the velocity of the Sun relative to the prefered frame is approximately given by

$$|v'| \approx 370 km/s$$

with the direction of the Sun to the CMB (prefered frame) (see Kogurt et al. [6])

$$(l'',b'') = (276° \pm 3°, 30° \pm 3°).$$

Since the trajectory of the spacecraft is approximately hyperbolic the velocities of the spacecraft $w_b'$ before and $w_a'$ after reaching the nearest point to the Earth are nearly constant but different from one another. The velocity $-v'$ of the non-prefered reference frame is also nearly constant. Therefore, equation (3.11) gives a constant value before and another constant value after reaching the nearest point to the Earth. In relation (3.11) the first expression of the right hand side may be compensated by the second one, i.e. there is no anomalous Earth flyby. The jump can be rewritten by virtue of $|w_b'| \approx |w_a'| \approx |w'|$ and (3.12) in the form

$$\left(\frac{w_b'}{c}, \frac{v'}{c}\right) - \left(\frac{w_a'}{c}, \frac{v'}{c}\right) \approx \left|\frac{w'}{c}\right|\left|\frac{v'}{c}\right|(\cos(w_b';v') - \cos(w_a';v')). \quad (3.13)$$

There exists another representation of the jump by the formula

$$\left(\frac{w_b'}{c}, \frac{v'}{c}\right) - \left(\frac{w_a'}{c}, \frac{v'}{c}\right) = \left(\frac{w_b'}{c} - \frac{w_a'}{c}, \frac{v'}{c}\right) = \left|\frac{w_b'}{c} - \frac{w_a'}{c}\right|\left|\frac{v'}{c}\right|\cos(w_b'-w_a';v') =$$

$$= 2\left|\frac{w'}{c}\right|\left|\frac{v'}{c}\right|\left|\sin((w_b';w_a')/2)\right|\cos(w_b'-w_a';v'). \quad (3.14)$$

Summarizing, formula (3.11) together with (3.12) and (3.13) implies that in the non-prefered reference frame of the Earth the anomalous Earth flyby of the spacecrafts does not exist but it is a consequence of Einstein's special relativity.

It is worth to mention a previous paper [7] where only the transverse Doppler effect is considered, i.e. the spacecraft during the nearest point to the Earth.

There exist several papers considering explanations of the anomalous Earth flyby, see e.g. [8-13].

Finally, the Pioneer anomaly shall be mentioned. The literature about possible explanations is very extensive. Here, I will only consider paper [3] where the Pioneers are studied in the universe. The total Doppler frequency shift is given by (see [3], formula (3.11))

$$\frac{d}{dt_e'}(v_{obs} - v(0,t_e)) = \frac{d}{dt_e'}\left(v_{obs} - \left(1 - \frac{v(t_e')}{c}\right)v_0\right) + \frac{d}{dt_e'}\left(\left(1 - \frac{v(t_e')}{c}\right)(1 - a(t_e))v_0\right). \quad (3.15)$$

Here, $v_{obs}$ is the observed frequency, $v(0,t_e)$ is the calculated frequency of the arriving photon in the universe and $v_0$ is the frequency emitted at present by the same atom at rest. The function $a(t)$ is the scaling factor describing the universe. The first expression of the right hand side of formula (3.15) is the anomalous frequency shift with the value

$$\dot{v} \approx 6 \cdot 10^{-9} \ Hz/s.$$

The used reference frequency is

$$v_0 = 2.29 \cdot 10^9 \ Hz.$$

The last expression in formula (3.15) follows by the universe and it gives the value

$$-H_0 v_0 \approx -5.3 \cdot 10^{-9}$$

where

$$H_0 \approx 70 \frac{km}{\sec Mpc} \approx 2.3 \cdot 10^{-18} \frac{1}{s}$$

is used.

Hence, the total frequency shift of the Pioneers in the universe is zero, i.e. there is no anomaly of the Pioneers. All these results with more details are derived in paper [3].

Conclusion: (a) The anomalous Earth flyby of spacecrafts does not arise if the non-prefered reference frame of the Sun (Earth) is used. (b) The Pioneer anomaly does not arise if the universe by studying the Pioneers in the solar system is not neglected.

Hence, the two anomalies are received by incomplete considerations and in reality they do not exist.